\documentclass[twocolumn]{aastex631}

\begin{document}

\title{Irradiation-Driven Formation of Supersoft X-ray Sources Following Classical Novae}

\author[0000-0002-1168-8221]{Weitao Zhao}
\affiliation{School of Physical, Henan normal university, Xinxiang 453007, China
\\e-mail: zhaoweitao@htu.edu.cn}
\affiliation{Center for Theoretical Physics, Henan Normal University, Xinxiang 453007, China}
\author{Xiangcun Meng}
\affiliation{Yunnan Observatories, Chinese Academy of Sciences, Kunming 650216, China\\e-mail: xiangcunmeng@ynao.ac.cn}
\affiliation{Key Laboratory for the Structure and Evolution of Celestial Objects, Chinese Academy of Sciences, Kunming 650216, China}
\affiliation{International Centre of Supernovae, Yunnan Key Laboratory, Kunming 650216, P. R. China}
\author{Yingzhen Cui}
\affiliation{Key Laboratory of Optical Astronomy, National Astronomical Observatories, Chinese Academy of Sciences, Beijing 100101, China}
\author{Yunlang Guo}
\affiliation{School of Astronomy and Space Science, Nanjing University, Nanjing 210023, China}
\affiliation{Key Laboratory of Modern Astronomy and Astrophysics, Nanjing University, Ministry of Education, Nanjing 210023, China}



\begin{abstract}

   Supersoft X-ray sources (SSSs) are characterized by persistent thermonuclear burning on the surfaces of white dwarfs (WDs). The standard model requires high mass transfer rates of $\sim 10^{-7}\, \rm M_{\odot}\,yr^{-1}$ from massive companions, presenting a theoretical impediment to the observed short-period SSSs, whose orbital periods imply low-mass donors theoretically incapable of sustaining such accretion. To resolve this paradox, we propose and demonstrate through detailed simulations that irradiative feedback following a classical nova (CN) eruption provides a natural formation channel. Through detailed binary evolution simulations with MESA, we reveal that sustained WD irradiation—initially from the outburst and subsequently from accretion luminosity—triggers significant and stable expansion of the low-mass companion. This, in turn, drives mass-transfer rates into the stable hydrogen-burning regime and sustains it beyond $10^4$ years after the initiation of hydrogen burning. This mechanism robustly explains the observed population of short-period SSSs. Moreover, when irradiation-driven mass transfer rate drops below the stable accretion rate, it may lead to the rapid accumulation of sufficient material on shorter time scales to trigger a recurrent nova outburst instead of SSS, thereby also offering an explanation for the origin of short-period recurrent novae.
   
\end{abstract}


\section{Introduction} \label{sec:intro}
Supersoft X-ray sources (SSSs) are characterized by high X-ray luminosities ($\sim 10^{36} - 10^{38} \rm erg/s$) and extremely soft spectra, with effective temperatures typically in the range of $10^{5} - 10^{6} $ K, corresponding to blackbody peaks at $\rm 15 - 85 eV$ \citep{Greiner1991, Alcock1996, vanTeeseling1996, Kahabka1997, Gansicke1998, Cowley2002}. These systems are widely interpreted as close binaries containing a white dwarf (WD) accreting from a non-degenerate companion (e.g., a main-sequence star or a red giant). The prevailing model posits that a high accretion rate ($\sim 10^{-7}\, \rm M_{\odot}\,yr^{-1}$) onto the WD leads to steady hydrogen burning on its surface, which produces the distinctive supersoft X-ray emission \citep[]{vandenHeuvel1992, Nomoto2007}. As this configuration mirrors the accretion scenario of the single-degenerate (SD) model for Type Ia supernovae (SNe Ia), SSSs are considered possible candidate progenitors for SNe Ia \citep[]{Nomoto1984, vandenHeuvel1992,Maoz2014}. 

The mass transfer rate from the donor star plays an important role in the accretion process onto WD. When the mass transfer rate exceeds the critical accretion rate to the WD, the excess material accumulates on the surface, forming an extended envelope \citep{Nomoto2007}. If the mass transfer rate falls below the stable accretion rate to the WD, the material accumulates on the surface of the WD, increasing temperature and pressure at the base of the envelope until hydrogen ignition triggers a nova outburst. Only when the mass transfer rate lies between the critical accretion rate and the stable burning rate, can hydrogen burn steadily on the WD surface, producing intense supersoft X-ray emission. This stable burning regime typically requires a massive main-sequence ($1.5 - 2.5 M_{\odot}$) companion star to supply a sufficiently high mass transfer rate ($\sim 10^{-7}\, \rm M_{\odot}\,yr^{-1}$) \citep{vandenHeuvel1992,Alcock1996,PH.2010,Nomoto2007}. However, several SSSs, such as SSS RX J0439.8 - 6809 \citep{Schmidtke1996b, vanTeeseling1997}, 1E 0035.4 - 7230 \citep{Schmidtke1996a}, and RX J0537.7 - 7034 \citep{Greiner2000}, exhibit extremely short binary orbital periods ($\sim $3 - 4 hours). This implies that their companion stars have very low masses ($< 0.5\, M_{\odot}$). For such low-mass main-sequence companions, their mass loss is mainly driven by angular momentum loss mechanisms, with a rate of only $10^{-10} - 10^{-9}\,{\rm M}_{\odot}\,{\rm yr}^{-1}$ — two to three orders of magnitude lower than the rate required by SSSs \citep{vandenHeuvel1992,Nomoto2007}. This paradox presents a significant challenge to standard binary evolution theory \citep{vanTeeseling1997,Greiner2000}, which predicts that such low-mass companions are incapable of supplying the required high mass-transfer rates.

A potential solution to this paradox may lie in the irradiation feedback mechanism, an idea with roots in the classical "hibernation hypothesis" for novae. This hypothesis was proposed to explain why some old novae remain bright for a century after outburst, positing that the phenomenon is caused by enhanced mass transfer triggered by irradiation of the red dwarf by radiation from the WD following the eruption \citep{Shara1986ApJ}. \citet{Kovetz1988} proposed that low-mass companions are heated after a nova outburst, leading to an increase in mass transfer rate by several orders of magnitude, although their model overlooked key physical processes such as accretion feedback. Subsequently, \citet{Ginzburg2021} advanced this theoretical foundation by developing more precise scaling relations and proposing a self-sustaining mechanism that offered deeper insights into the system dynamics. However, their models were constrained by analytical approaches, lacking comprehensive long-term, self-consistent binary evolution simulations and failing to fully incorporate critical factors such as irradiation efficiency and the detailed structural evolution of the companion star. In contrast to these irradiation-based models, the wind-driven evolution proposed by \citet{vanTeeseling1998A&A} could potentially provide such a high mass transfer rate through a different mechanism, though its triggering and long-term sustainability remain unclear.  This further underscores the need for the kind of self-consistent simulations we present here. Recently, \citet{Zhao2022,Zhao2024AA} introduced the mechanism of X-ray irradiation of the companion by the WD radiation in SSSs, and conducted detailed irradiated companion simulations using MESA, successfully explaining the quasi-periodically optical light curve.

In this study, we conduct the first detailed, long-term binary evolution simulations to rigorously test the irradiation feedback mechanism as the formation channel for short-period long-lived SSSs. Using the Modules for Experiments in Stellar Astrophysics (MESA) code \citep[version 10398;][]{Paxton2011,Paxton2013,Paxton2015,Paxton2018,Paxton2019}, we model the post-nova evolution and show that irradiation can dramatically increase mass transfer from low-mass companions. We systematically explore the parameter space and resulting mass distribution enabled by this channel. In Section~\ref{sec:Method}, we describe the setup and methodology, and presents the simulation results in Section~\ref{sec:results}. In Section~\ref{sec:Discussion}, we discuss the implications of our findings, and summarize our main conclusions in Section~\ref{sec:summary}.

\section{Method} \label{sec:Method}

In this study, we used MESA to investigate the response of a low-mass main-sequence (MS) companion star to irradiation following a nova eruption. The initial condition is a companion star on the zero-age main sequence (ZAMS), filling its Roche lobe. Here, the orbital period is determined by the mass of the companion star and the WD mass. Our simulations considered WD with masses ranging from 0.6 to 1.2 $\rm M_{\odot}$, which were treated as point masses. These WDs were accompanied by MS companions with masses between 0.2 and 2.0 $\rm M_{\odot}$. The companion star models were computed with a mixing-length parameter of 2.0 and a metallicity of $Z = 0.02$ \footnote{The inlists for our models are available at \href{https://doi.org/10.5281/zenodo.17935280}{https://doi.org/10.5281/zenodo.17935280}}.  

In classical nova (CN) systems, the WD accretes material from the companion star, which loses material due to the loss of the orbital angular momentum at a rate typically on the order of $10^{-10} - 10^{-9}\,{\rm M}_{\odot}\,{\rm yr}^{-1}$. When a nova eruption occurs, the nova luminosity (and possible supersoft X-ray luminosity due to residual envelope burning) was applied to irradiate the companion star for a duration of 0.1 years. Subsequently, the dominant irradiation source transitions to the accretion luminosity, given by \(L_{\rm acc} = G M_{\rm WD} \dot{M}_{\rm acc} / R_{\rm WD}\), where \(M_{\rm WD}\) is the WD mass, $\dot{M}_{acc}$ is the accretion rate to the WD, and $R_{\rm WD}$ is the radius of the WD. The effective irradiating flux intercepted by the companion is calculated as $L_\mathrm{irr} = \eta (R_{2}/2a)^2 L$, where $L$, ${\eta}$, $R_2$, and $a$ represent either the post-nova or accretion luminosity, the irradiation efficiency, the radius of the companion star, and the orbital separation, respectively \citep[]{Milgrom1978, Mason1989, Ritter2000}. Here, we treat the WD as a point mass to focus on the core feedback effects of irradiation on the structure of the companion and mass transfer rate. To ensure physical accuracy in calculating the accretion luminosity, we employ empirical relations for the WD radius, where $R_{\rm WD}$ is taken to be the photospheric radius of a WD undergoing stable hydrogen burning \citep{Nomoto2007}. Some studies suggest that the irradiation efficiency is at least 0.1 \citep{Hameury1993,Ritter1994,King1995,Ritter2000}, and here we adopt the minimum value of 0.1, which may be somewhat underestimated.

The irradiation energy is added to the surface of the companion star under the simplifying assumption of spherical symmetry \citep{PH.1991,Hameury1993,Zhao2024AA}, with the penetration depth determined by the radiative attenuation law, \(I/I_0 = e^{-\tau}\) \citep{Kovetz1988,PH.1991}, where $\tau$, $I_{\rm 0}$, and $I$ are the optical depth, initial radiation intensity incident on surface of the companion star, and radiation intensity at a depth within the surface of the companion star, respectively. When the accumulated mass on the WD reached a critical value (dependent on the WD mass) and triggered a hydrogen ignition event, the resulting supersoft X-ray luminosity was then used to continuously irradiate the companion, and the subsequent the long-term evolution of the companion star under the irradiation feedback mechanism was tracked. The mass transfer rate was calculated using the model described by \citet{Kolb1990}. We examined different WD and companion star masses, as well as various orbital periods, to map out the parameter space for nova-induced SSSs. 

\section{Results} 
\label{sec:results}

 \begin{figure}
        \centering
        \includegraphics[width=0.9\columnwidth]{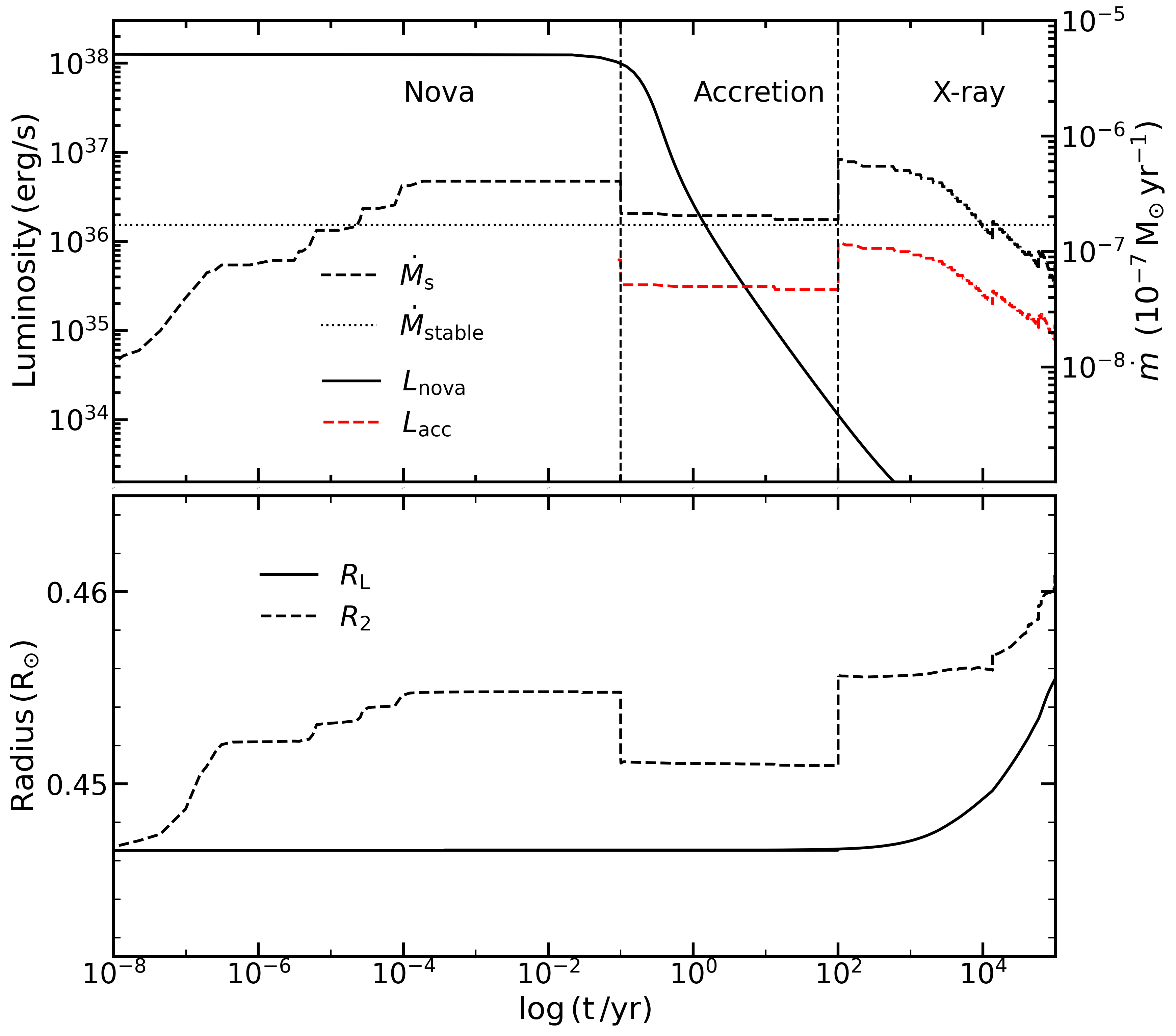}
        \caption{Irradiation-induced changes in mass transfer rate and companion star radius following nova outburst in our model. The WD mass, mass of the companion star, the orbital period, and irradiation efficiency are 1.0 ${\rm M}_{\odot}$, 0.5 ${\rm M}_{\odot}$, 0.154 days, and 0.1, respectively. Upper Panel: The black solid line represents the modeled luminosity during the nova outburst for the WD mass of 1.0 ${\rm M}_{\odot}$. The black dashed line represents the mass transfer rate modulated by irradiation of the companion star. The red dashed line shows the accretion luminosity resulting from material accreted onto the WD in the aftermath of the outburst. The horizontal black dotted line indicates the stable accretion rate for a WD of 1.0 ${\rm M}_{\odot}$. Lower Panel: The black solid and dashed lines represent the stellar and the Roche lobe radii of the companion star, respectively.}
        \label{fig:1} 
   \end{figure}
   
\subsection{Irradiation Feedback on the Companion}

Figure \ref{fig:1} illustrates the evolutionary response of the nova system following an outburst in our simulation for an irradiation efficiency of 0.1. The WD accretes material at a rate of $\sim 10^{-10}\,{\rm M}_{\odot}\,{\rm yr}^{-1}$ and continues to accumulate material until a critical mass is reached, triggering a nova outburst. During nova outburst phase, the irradiation by the nova luminosity ($\sim 10^{38}$), drives the mass transfer rate to increase rapidly, reaching $\sim 4 \times 10^{-7}\,{\rm M}_{\odot}\,{\rm yr}^{-1}$. Concurrently, radiative heating causes the radius of the companion star to expand from $0.446\, {\rm R}_{\odot}$ to $0.455 \,{\rm R}_{\odot}$. Throughout this phase, the presence of nova ejecta effectively prevents accretion onto the WD, resulting in negligible accretion luminosity. After the outburst subsides, the system luminosity declines to a quiescent level, allowing accretion onto the WD to resume. The resulting accretion luminosity $L_{\rm acc}$ then irradiates the companion, establishing a self-sustaining feedback loop: as the mass transfer rate $\dot{M}$ decreases, $L_{\rm acc}$ declines until the system reaches an equilibrium near $\dot{M} \sim 2 \times 10^{-7}\,{\rm M}_{\odot}\,{\rm yr}^{-1}$, accompanied by a contraction of the companion radius to $0.45 \,{\rm R}_{\odot}$. This near-equilibrium state persists until the accumulated envelope mass on the WD triggers stable hydrogen burning, transitioning the system into a supersoft X-ray phase that lasts approximately 100 years, i.e., self-sustained mass transfer rate. This extended duration allows the WD of 1.0 $\rm M_{\odot}$ to accumulate sufficient mass to initiate hydrogen burning, and we hypothesize that the WD then undergoes sustained hydrogen burning, transitioning the system into a SSS phase. During the SSS phase, irradiation by supersoft X-rays becomes the dominant heating mechanism, elevating $\dot{M}$ to a peak value near $\sim 6.2 \times 10^{-7}\,{\rm M}_{\odot}\,{\rm yr}^{-1}$. After the mass transfer rate peaks, the companion star, due to ongoing rapid mass loss, gradually contracts, leading to a decrease in the mass transfer rate. Then the mass transfer rate remains above $\dot{M}$ ($> 10^{-7}\,{\rm M}_{\odot}\,{\rm yr}^{-1}$) for over $ 5 \times10^{4}$ years, while the radius of the companion star increases progressively and continuously exceeds its Roche lobe radius. 

Our results indicate that irradiation from the post-nova system—including that produced by accretion luminosity—can cause the companion star to expand beyond its Roche lobe, leading to a substantial and sustained increase in the mass transfer rate. The irradiation feedback increased the mass transfer rate of the system by at least three orders of magnitude, from an extremely low level (only $\sim 10^{-10}\,{\rm M}_{\odot}\,{\rm yr}^{-1}$) that could not sustain stable burning to the critical region where stable burning is possible. This leap in scale cannot be achieved by traditional evolutionary mechanisms. The formation of an SSS is found to be highly dependent on system parameters, meaning not all nova systems follow this evolutionary path. Given this parameter sensitivity, a systematic investigation of the critical conditions for SSS formation is essential. In the following section, we will explore the parameter space conducive to SSS emergence, with particular focus on the binary orbital period, the WD mass, and the mass of the companion star.

\subsection{Parameter Space Analysis}
 \begin{figure}
        \centering
        \includegraphics[width=0.9\columnwidth]{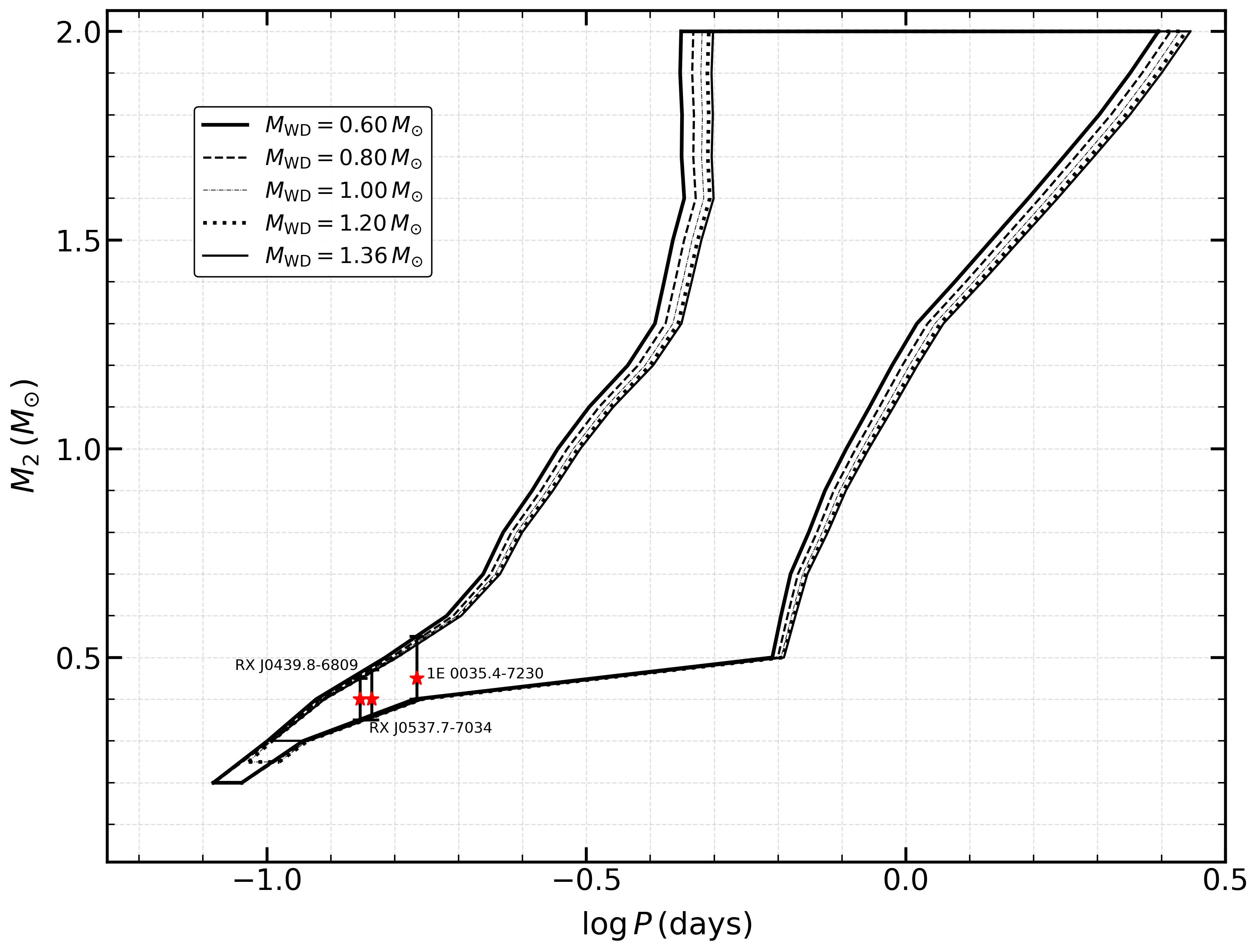}
        \caption{Regions in the initial orbital period-secondary mass plane (${\rm log} P$, $M_{2}$) for WD binaries that produce SSSs for initial WD masses of 0.60,\, 0.80, 1.0, 1.2, and 1.36 $\rm M_{\odot}$. Here, the irradiation efficiency is 1. The red symbols represent observed short period SSSs in this model, with the examples RX J0439.8–6809 \citep{Schmidtke1996b,vanTeeseling1997}, 1E 0035.4–7230 \citep{Schmidtke1996a}, and RX J0537.7–7034 \citep{Greiner2000} corresponding to their potential locations and inferred companion mass ranges.}
        \label{fig:2} 
   \end{figure}

Figure \ref{fig:2} shows the parameter contours for successful SSS formation following classic nova outbursts. Here, to demonstrate the maximum parameter range, we assume a radiation efficiency of 1, although this is somewhat overestimated. The left boundary of the contours corresponds to systems in which Roche-lobe overflow begins while the secondary star is still on the ZAMS. The right boundary reflects the evolutionary stage at which the systems will experience dynamically unstable mass transfer at the base of the red-giant branch (RGB). The upper boundary is set by the companion mass above which a system can evolve into an SSS through standard (non-irradiated) evolution. Specifically, when the companion mass exceeds 2.0 $\rm M_{\odot}$, an SSS can be formed solely through the evolution of the companion without irradiation. The lower boundary, in contrast, represents the minimum companion mass required for irradiation-driven SSS formation. Below this threshold, the accretion luminosity onto the WD is insufficient to maintain a mass transfer rate high enough to trigger stable hydrogen burning, preventing the transition to an SSS phase. If the initial binary parameters fall within the SSS formation region shown, the system is expected to undergo binary evolution leading to an SSS phase. Here, when the companion mass is greater than 1.5 $\rm M_{\odot}$, the triggering mechanism of the SSS is not necessarily a nova outburst, but more likely mass transfer on a thermal timescale. However, at least this mechanism can enhance the mass transfer rate.

The existence of stable nuclear burning on WD in SSSs requires mass transfer rates from a companion star on the order of ($\sim 10^{-7}\,{\rm M}_{\odot}\,{\rm yr}^{-1}$) \citep{vandenHeuvel1992,Nomoto2007,PH.2010}. Standard binary evolution theory posits that such high, sustained accretion rates are typically achievable only with a donor star that is massive enough (at least 1.5 $\rm M_{\odot}$) \citep{PH.2010}, a scenario which corresponds to an orbital period exceeding approximately 0.4 days. This theoretical framework, however, is challenged by the observed properties of several SSSs. The systems RX J0439.8 - 6809 \citep{Schmidtke1996b, vanTeeseling1997} (orbital period $P_{\rm orb} = 0.139\,d$), 1E 0035.4 - 7230 ($P_{\rm orb} = 0.171\, d$)\citep{Schmidtke1996a}, and RX J0537.7 - 7034 ($P_{\rm orb} = 0.146 \,d$) \citep{Greiner2000} (these red symbols in Fig.\ref{fig:2}) all reside in compact binaries whose short periods imply low-mass companions, even lower 0.5 $\rm M_{\odot}$. Such low-mass donors are theoretically incapable of supplying the high $\dot{M}$ necessary for steady hydrogen burning, presenting a clear contradiction between observation and canonical binary evolution models.

Most importantly, the currently observed short-period SSS sources are all precisely located within our predicted SSS formation region (Fig.\ref{fig:2}). This strongly indicates that irradiation feedback is the missing link connecting the theoretically impossible with the observationally present, providing a unified and self-consistent explanation for the entire sample. The efficacy of this irradiation-driven mass transfer paradigm extends beyond SSSs; it also provides a consistent explanation for the high accretion rates observed in short-period recurrent novae, with more derivations given in Discussion \ref{sec:Discussion}.

\section{Discussion}
\label{sec:Discussion}
 \begin{figure}
        \centering
        \includegraphics[width=0.9\columnwidth]{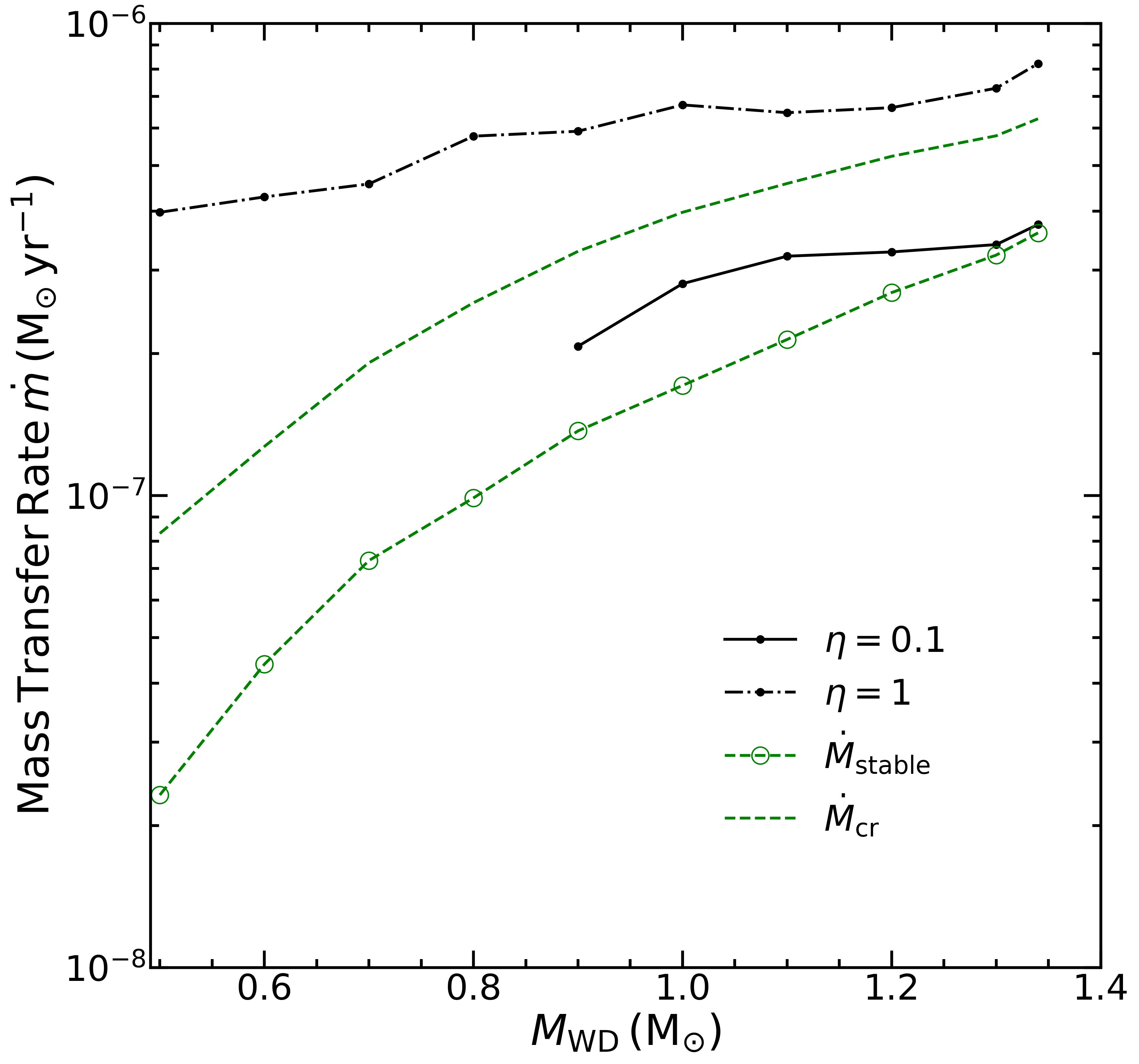}
        \caption{Possible properties of hydrogen-burning shells onto the accreting WDs in the plane of the WD mass, $M_\mathrm { WD}$, and the mass transfer rate, $\dot{M}$. The green dashed and solid lines represent $\dot{M}_\mathrm { stable}$ and $\dot{M}_\mathrm { cr}$, respectively. The black dashed and solid lines represent the irradiation efficient $\eta = 1$ and $\eta = 0.1$, respectively.}
        \label{fig:3} 
   \end{figure}
    \begin{figure}
        \centering
        \includegraphics[width=0.9\columnwidth]{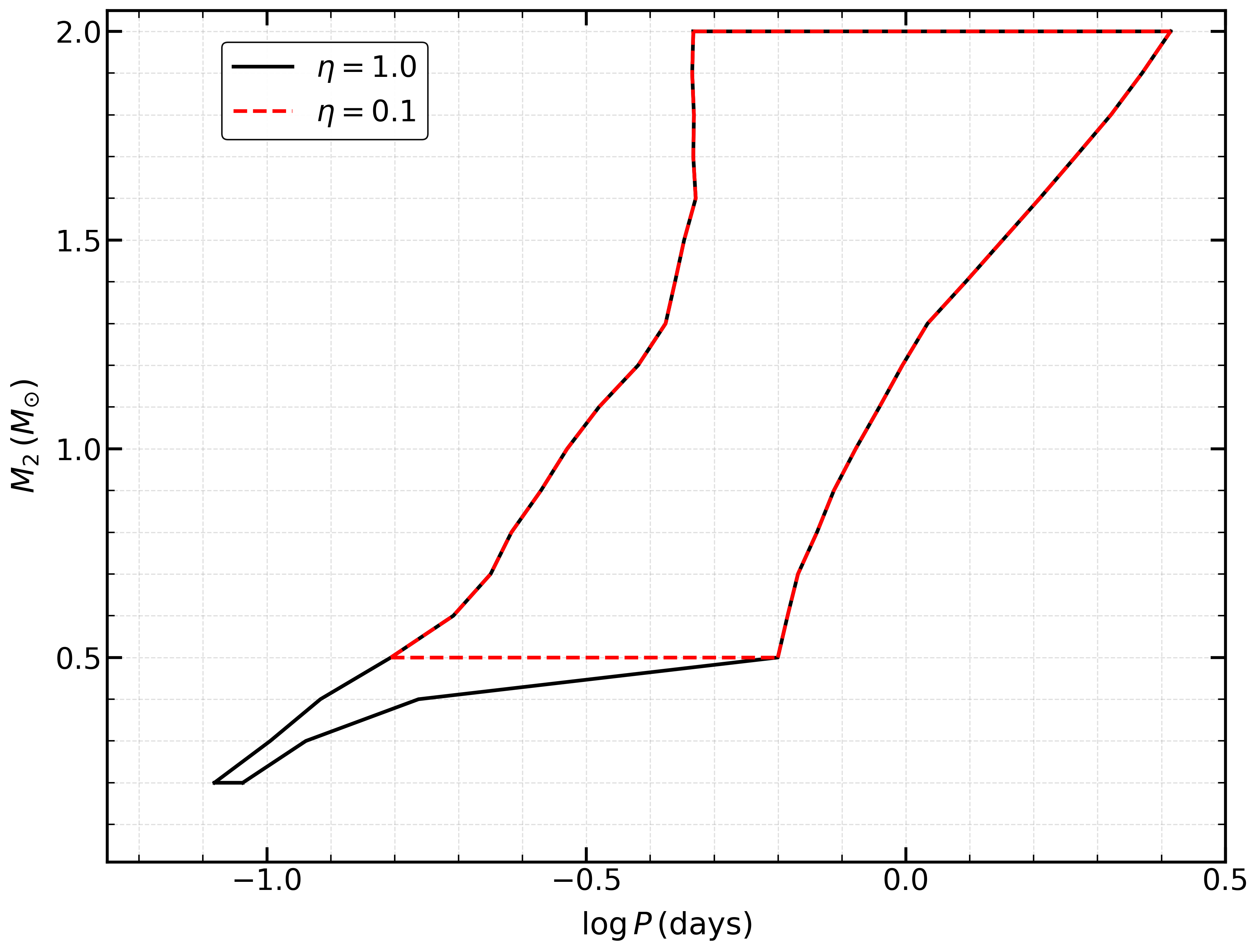}
        \caption{Similar to Figure \ref{fig:2} but for different irradiation efficiently, and the initial WD masses was set be 0.9$\rm M_{\odot}$. }
        \label{fig:4} 
   \end{figure}
   
In this study, we have investigated a binary evolution mechanism where the companion star is irradiated by both the persistent accretion luminosity and the intense outburst luminosity following a nova eruption. Using the MESA code, we conducted simulations that model the WD as a mass point and self-consistently treat the irradiation feedback on the donor star. Our models demonstrate that irradiation can trigger thermally stable mass transfer, causing the system to evolve into a SSS within a timescale of several years to several centuries post-eruption, depending on the WD mass. This SSS phase can subsequently be sustained for over $10^4$ years. These findings significantly expand the theoretical parameter space for SSS formation and provide a robust explanation for the existence of short-period SSS systems.

The irradiation of the companion star by the WD may be partially blocked by the accretion disk, though we expect this obscuration to be less significant than in neutron star systems. We conservatively assume that at least $10$ percent of the energy of the irradiating flux reaches the hemisphere of the companion facing the WD \citep{Hameury1993}. To assess the sensitivity of our results to this key parameter, we explore how the irradiation efficiency $\eta$ influences both the self-sustained mass transfer rate and the resulting parameter space for SSS formation. As shown in Figure\ref{fig:3}, for a high efficiency ($\eta = 1$), the self-sustained mass transfer rate surpasses the critical accretion rate for steady hydrogen burning across a wide range of system parameters. In contrast, for a lower efficiency ($\eta = 0.1$), this condition is only met for WDs more massive than 0.9 $\rm M_{\odot}$. 

Following a nova eruption, a transient supersoft X-ray phase is commonly observed, typically lasting from several months to years \citep{Hachisu2010ApJ}. The duration of this initial phase is governed by the mass of the residual hydrogen envelope on the WD, which strongly depends on the WD mass, with more massive WDs exhibiting shorter SSS phases due to higher nuclear burning rates. Our simulations predict that after the outburst, the nova system will transition into a SSS over a longer timescale (far exceeding the timescale of the residual hydrogen envelope burning), which will result in an expected spatial coincidence between CNe and SSS. This theoretical framework finds observational validation in Nova Ophiuchi 1938, which was initially detected as a CN in 1938 \citep{Hogg1964,Margon1991} and subsequently identified as a persistent SSS in 2023 \citep{Zhao2024MNRAS}. Here, the interval from the nova outburst to the discovery of SSS reached 85 years. Furthermore, systematic monitoring in M31 has identified several novae with SSS emission initiating 3–9 years post-outburst \citep{Stiele2010, Pietsch2010,Henze2010}. In addition, if the nova outburst does not disrupt the accretion process, our model can also account for the prolonged SSS phase following the nova eruption. For example, V723 Cassiopeiae exhibited an SSS phase lasting fifteen years after its nova outburst \citep{Ness2008}, while 1RXS J050526.3–684628 maintained this phase for over thirty years \citep{Vasilopoulos2020}. Future long-term monitoring is crucial to constraining the parameters of these systems and testing our model. 

Our model not only explains SSSs, but is also effective in accounting for another class of observational phenomena — short period recurrent novae (RNe), such as IM Nor \citep{Elliot1972} and T Pyx \citep{Shahbaz1997}. When the irradiation efficiency is lower than 0.1 (even lower), the self-sustained mass transfer rate may lie slightly below the steady hydrogen burning rate, the system is more likely to evolve into a RN. In this regime, the WD undergoes repeated thermonuclear eruptions as it accretes material at a high but lower than stable hydrogen burning rate. The model reveals a profound connection: CNe, RNe, and SSSs may represent different manifestations of the same underlying physical process, e.g., irradiation-driven mass transfer—manifesting differently under varying system parameters such as irradiation efficiency and WD mass. When the irradiation efficiency is sufficiently high, the system evolves into an SSS; at slightly lower efficiencies, it manifests as an RN. This provides a novel and unified theoretical framework for understanding these closely related astrophysical phenomena. 

\section{Summary}\label{sec:summary}

This study investigates how irradiation from both the nova eruption and the subsequent accretion luminosity plays the pivotal role in driving the evolution of CN systems. We demonstrate that the intense radiation during a nova outburst heats and expands the atmosphere of the companion star, triggering a sharp increase in the mass transfer rate. Following the eruption, persistent irradiation by accretion luminosity can maintain the mass transfer rate at an elevated, stable level. When this sustained rate exceeds the stable hydrogen burning rate to the WD for a sufficiently long duration (on the order of a hundred years, depending on WD mass), the system transitions into a SSS, which may persist for over $10^4$ years.

This study establishes that irradiation feedback following nova outbursts is the key physical mechanism responsible for forming short-period SSSs, such as RX J0439.8-6809\citep{Schmidtke1996b,vanTeeseling1997}, 1E 0035.4–7230 \citep{Schmidtke1996a}, and RX J0537.7–7034 \citep{Greiner2000}, thereby resolving a long-standing theoretical contradiction in the field. Furthermore, the model predicts sustained supersoft X-ray emission at the sites of past novae—a phenomenon with growing but still limited observational support, as in the case of Nova Ophiuchi 1938. In systems where the irradiation-sustained mass transfer rate is slightly lower than $\dot{M}_{stable}$, the binary may enter a recurrent nova phase, offering a natural explanation for frequently erupting systems such as IM Normae \citep{Elliot1972} and T Pyx \citep{Shahbaz1997}.

\begin{acknowledgments}
      We acknowledge the anonymous referee for the valuable comments that helped improve this paper. This work is supported by the National Natural Science Foundation of China (Nos. 12333008 and 12288102, 12403035) and National Key R\&D Program of China (No. 2021YFA1600403). X.M. acknowledges support from the Strategic Priority Research Program of the Chinese Academy of Sciences (grant Nos. XDB1160303, XDB1160000), the National Science Foundation of China and National Key R\&D Program of China (No. 2021YFA1600403), Yunnan Fundamental Research Projects (NOs. 202401BC070007 and 202201BC070003), International Centre of Supernovae, Yunnan Key Laboratory (No. 202302AN360001), the Yunnan Revitalization Talent Support Program-Science \& Technology Champion Project (NO. 202305AB350003), and the China Manned Space Program with grant (No. CMS-CSST-2025-A13). The Postdoctoral Fellowship Program of CPSF under Grant (NO. GZB20240307), the China Postdoctoral Science Foundation under Grant (NOs. 2024M751375 and 2024T170393), and the Jiangsu Funding Programme for Excellent Postdoctoral Talent under Grant (NO. 2024ZB705).
\end{acknowledgments}

\appendix

\bibliography{Nova_SSS}{}
\bibliographystyle{aasjournal}



\end{document}